# Atomic Control of Strain in Freestanding Graphene

P. Xu,[1] Yurong Yang,[1,2] S.D. Barber,[1] M.L. Ackerman,[1] J.K. Schoelz,[1] D. Qi,[1] I.A. Kornev,[3] Lifeng Dong,[4,5] L. Bellaiche,[1] S. Barraza-Lopez,[1] and P.M. Thibado[1*]

[1]*Department of Physics, The University of Arkansas, Fayetteville, Arkansas 72701, USA*

[2]*Physics Department, Nanjing University of Aeronautics and Astronautics, Nanjing 210016, China*

[3]*Ecole Centrale Paris,CNRS-UMR8580, Grande Voie des Vignes, 92295, France*

[4]*College of Materials Science and Engineering, Qingdao University of Science and Technology, Qingdao 266042, China*

[5]*Department of Physics, Astronomy, and Materials Science, Missouri State University, Springfield, Missouri 65897, USA*

In this study, we describe a new experimental approach based on constant-current scanning tunneling spectroscopy to controllably and reversibly pull freestanding graphene membranes up to 35 nm from their equilibrium height. In addition, we present scanning tunneling microscopy (STM) images of freestanding graphene membranes with atomic resolution. Atomic-scale corrugation amplitudes 20 times larger than the STM electronic corrugation for graphene on a substrate were observed. The freestanding graphene membrane responds to a local attractive force created at the STM tip as a highly-conductive yet flexible grounding plane with an elastic restoring force. We indicate possible applications of our method in the controlled creation of pseudo-magnetic fields by strain on single-layer graphene.

PACS numbers: 68.65.Pq, 68.37.Ef, 31.15.A-, 31.15.aq

The extraordinary properties of graphene, all exhibited within a single plane of carbon atoms, continue to drive a frenzy of research [1]. In most graphene studies, samples are on a substrate, which degrades the intrinsic mobility of graphene [2]. The mechanisms behind this degradation include local effects, such as charged-impurity scattering [3], and nonlocal phenomena, such as phonon scattering [4]. Scanning tunneling microscopy (STM) and spectroscopy (STS) [5,6] reveal that charge-donating substrate impurities create charge puddles in supported graphene. The numerous limitations associated with examining graphene on substrates have led researchers to suspend graphene over holes [7,8] to better study its intrinsic properties. These efforts have been rewarded with many important breakthroughs, including the measurement of its record-breaking ballistic carrier mobility [9], thermal conductivity [10], and the fractional quantum Hall effect [11]. Freestanding graphene has also provided a way to probe the material's intrinsic tensile strength [12, 13]. Atomic force microscopy, combined with other techniques, has been utilized to measure its effective spring constant, resonance frequency (in the megahertz range) [14], Young's modulus, self-tension, and the breaking strength of single- and multiple-layer graphene [15-17]. More recently, STM has been used to create mini-membranes by locally lifting graphene from the substrate [18]. In addition, through the distortion of the two-dimensional plane with strain, the properties of charge carriers in graphene have been found to change dramatically as gauge fields (pseudo-magnetic and deformation potential) are created [19-22].

Researchers using transmission electron microscopy pioneered the efforts of imaging freestanding graphene, providing insight into the existence of ripples [23] and revealing point defects and ring defects, as well as edge reconstructions [24]. In this Rapid Communication, we describe a new experimental approach using constant-current STS to introduce strain in a controlled way into the freestanding graphene. We also successfully obtain atomic-resolution STM images and document vertical corrugations *(d)* that are 20 times larger than the expected electronic corrugation $(d_e)$ [25] due to strain induced movement *(u, where $d = d_e + u$)*.

Experimental constant-current STS and STM measurements were obtained using an Omicron ultrahigh-vacuum low-temperature STM operated at room temperature. Two different samples were examined. The first was a 2000-mesh, ultrafine grid with a square lattice of holes with side $2L$ = 7.5 μm and copper bar supports 5 μm wide, upon which graphene had been transferred [26]. The grids were mounted on flat tantalum STM sample plates using silver paint. STM data was acquired on freestanding graphene suspended over the holes as well as attached to the copper bars. We also used a piece of highly-oriented pyrolytic graphite (HOPG), the top layers of which had been removed with adhesive tape to expose a fresh surface only for comparison with our freestanding graphene data. Samples were imaged using STM tips manufactured in-house by electrochemically etching polycrystalline tungsten wire via a differential-cutoff lamella method. After etching, the tips were gently rinsed with distilled water, briefly dipped in a concentrated hydrofluoric acid solution to remove surface oxides, and transferred through a load-lock into the STM chamber. The morphology of graphene layers grown on the copper grid was also examined using an FEI Quanta 200 field emission scanning electron microscope (SEM).

An SEM image of the copper grid with the graphene membranes is shown in Fig. 1(a). Notice that some of the holes in the copper mesh are fully covered by graphene, while others are partially covered. We estimate that graphene covers more than 90% of the grid. Constant-current scanning tunneling spectroscopy (CC-STS) data was acquired for both freestanding graphene and graphene on copper as shown in Fig. 1(b). The height ($d$) of the STM tip required to maintain a constant current is shown, as the bias between the tip and the grounded sample is varied. Constant-current data (feedback on) was acquired instead of the typical constant-height data (feedback off), because in the latter case the sample moves and either crashes into the tip or falls out of tunneling when the voltage is increased or decreased, respectively. The inset shows the actual measured current versus voltage on a log scale. Notice that constant current is achieved, indicating that the sample and tip always maintain a close separation to sustain electron tunneling. For the highest set-point current, the freestanding graphene membrane raises and *follows the tip in registry* by about $d = 30$ nm when the voltage reaches 3 V. The result is almost fully reversible as the graphene drops by about 35 nm when the voltage is ramped back down. Graphene can be held at any height in between by simply changing the voltage accordingly. The set-point current plays a similar role in the maximum displacement achieved. A tenfold reduction in current reduces the maximum height achieved by the freestanding graphene by about a factor of two. Therefore, a combination of set-point current and bias (for fine tuning) can be employed to set the height ($d$) of the graphene membrane at will. The CC-STS technique is ideal for quantifying the movement of the freestanding graphene since the voltage is incrementally changed in small amounts (~10 mV step size) and the acquisition waits for a long time (~3 ms) so that the feedback can stabilize before the current and new vertical position of the STM tip is recorded.

Constant-current STM images were also acquired from these samples. An STM image of graphene on copper is displayed in Fig. 2(a). The honeycomb structure is visible but somewhat obscured by the harsh morphology of the copper substrate, which gives an overall texture to the topography. The upper right inset shows an atomic resolution image from the central section of Fig. 2(a) around a single honeycomb ring, magnified 2 times, and displayed with a compressed color scale. Below the image is a height cross-section line profile extracted from the center of the STM image, showing an atomic-scale corrugation ($d_e$) of about 0.05 nm [25]. The vertical scale is fixed at 1 nm for all the line profiles for easy comparison. An STM image of the freestanding graphene is displayed in Fig. 2(b). The honeycomb structure is still visible but much less distinct due to the lateral movement of the freestanding graphene membrane (described further in sections to follow). This graphene image also shows an overall curvature to the topography. Here it is highest and nearly constant along the diagonal line from lower left to upper right. The most remarkable feature of this data is that the underlying atomic lattice is still visible even with an overall black-to-white color scale of 4 nm. The inset shows the local atomic structure with apparent distortions. The atomic lattice is visible because the height change ($d$) across the honeycomb is an astonishing full nanometer (see line profile below image), 20 times greater than the expected electronic corrugation ($d_e$) for graphene on a substrate [25]. For comparison, a high-quality STM image of graphene on graphite [27,28] is presented in Fig. 2(c). From the line profile below the image, the corrugation amplitude is again about 20 times smaller than that of the freestanding graphene membrane. Note the hexagons in these STM images confirm that it is *single-layer* graphene [29].

We now present a number of theoretical results which point to the significance of the controlled creation of strain fields (i.e., elastic distortions) on graphene by CC-STS. Given that the tip interacts with the membrane at a point, we first took a small graphene supercell with fifty atoms, width $2L \approx 2$ nm, and pulled the center atom, leaving the corner atoms fixed. The central atom was gradually displaced an amount $d$, through a series of 0.03 nm increments, up to a maximum perpendicular displacement of 0.30 nm from the plane. At each displacement, the other atoms were allowed to relax to the energy minimum. The atomic relaxation was carried within the local-density approximation to density-functional theory (DFT) with projector-augmented wave potentials [30] as implemented in the plane-wave basis set VASP code [31]. The restoring elastic force was obtained from Hellman-Feynmann theorem as the derivative of the energy data versus displacement and displayed in Fig. 3(a) [inset shows a schematic of the supercell in side view, while Fig. 3(b) shows the strained model in a tiled view]. The linear fit shown for distortions with $d/L < 0.1$ produced a spring constant of about 20 nN per nm. With the reported maximum vertical distortion, and assuming the STM tip was near the center of the suspended membrane, we obtained a $d/L$ = (35 nm)/(3.25 μm) ~0.01 from our data in Fig. 1, well within the linear region. This normalized displacement yields a restoring force of about 0.25 nN from Fig. 3(a).

We next estimated the attractive force between the STM tip and the graphene membrane with a simple calculation in which we computed the charge rearrangement in graphene needed to screen a unit point charge placed 0.1 nm above the graphene plane, using the self-consistent Hückel method. We used a 20 nm × 20 nm square patch of graphene containing more than 15,000 π-electrons for this purpose. The Coulomb force was computed from the self-consistent charge distribution. The van der Waals contributions were not considered in our calculation, because the charge accumulation component of the attractive force was sufficient to provide us with a qualitative picture with good insight into the origins of the atomic-scale corrugation.

Indeed, the electrostatic force varies periodically across graphene as shown in Fig. 3(c). Moving parallel to the graphene surface along a straight line toward one of the nearest carbon atoms, the force oscillates between a minimum (0.15 nN) when the tip is directly above a hole and a maximum (0.25 nN) directly above an atom. Thus, the attractive electrostatic force between the tip and the graphene oscillates in perfect registry with graphene's atomic spacing. When the test charge was directly above a point halfway between a hole and an atom, a lateral force was found to exist (not shown). The magnitude of this lateral force is significant, at more than 10% of the vertical force. It draws the graphene sheet sideways, causing the STM images of freestanding graphene to blur. Given that the force is not constant during the scan, the STM tip must retract or approach the graphene membrane as it scans the surface. The freestanding graphene follows the tip, as described previously; hence, the atomic-scale corrugation ($d$) is much larger than that observed when graphene is on a substrate ($d_e$).

Knowing from our experimental data and our DFT calculation that the membrane strain is within the linear region, we used the analytical expressions for the deformation of a membrane from Ref. [32], along with the expressions for the pseudo-magnetic field in Ref. [19] and the deformation potential in Ref. [33] to estimate the pseudo-magnetic field created by the induced strain ($d/L$) in our setup. Given that $d/L$ is small, in our calculation, we further assumed a circular profile at the edges of the graphene membrane where it contacted the copper. As a result, the pseudo-magnetic field has the following form: $B(r,\theta) = |B(r)|\cos(3\theta)$ [19]. We plotted the radial average of $|B(r)|$ between $r = 0$ and $r = 5$ nm vs. normalized displacement at $r = 0$ in Fig. 3(d). For a normalized displacement consistent with our results ($d/L = 0.01$), the $|B(r)|$ is about 25 T beneath the STM tip.

Historically, constant-current STS data like that shown in Fig. 1(b) was used to measure the local work function $\phi$ using the following formula: $\phi(eV) = 0.95[\Delta \ln V/\Delta d(\text{Å})]^2$, where V is the bias voltage and $d$ is the tip height [34]. (Note that because the measurement of the local work function is such a technologically critical parameter, STS was quickly superseded with the three-terminal ballistic electron emission microscopy method [35]). Applying this formula to our graphene on copper data yielded a work function of about 1 eV, which was smaller than the expected value of about 4 eV, but reasonable for STS [36,37]. Applying this formula to our freestanding graphene data, we found an "effective work function" of about 10 meV, or 100 times smaller, but here the graphene membrane is free to move and follows the retracting tip. During a standard CC-STS measurement on a substrate, the sample does not move.

The origin of the movement of the freestanding graphene membrane can be explained by a number of factors working in tandem. First, as the voltage increases, the current will increase due to an increasing tunneling probability. With the feedback turned on, the tip pulls away from the surface. However, the tip need only move about 0.1 nm to decrease the current by a factor of ten, so its movement will always be less than a few tenths of a nanometer [25]. More relevant to this study, the increased bias voltage increases the electrostatic force on the freestanding graphene membrane, causing it to move toward the tip as illustrated in Fig. 1(c). To maintain a constant current, the tip pulls away from the approaching graphene membrane. The membrane continues to chase the tip until an elastic restoring force grows and eventually equals the electrostatic force, as shown in the right side of Fig. 1(c). However, because there is a random offset in the height of a few nanometers in the CC-STS data shown in Fig. 1(b), this may be evidence of different large-scale, low-energy configurations, or ripple textures, of the graphene membrane [23,38,39]. This may also be the root cause of the two different linear regions in the spectra. The first linear region is soft due to removing ripples and the second region is hard due to straining the lattice. In this way, the CC-STS measurement on freestanding graphene is a tool by which we can learn about the elastic properties and the induced electronic properties of freestanding graphene. It is also interesting that for a fixed bias voltage (e.g., 3 V), the height $d$ increases as the current setpoint increases, counter to tunneling theory but consistent with our electrostatic attraction model.

Similar to the STS data, the STM data shown in Fig. 2(b) has two features that contribute to its atomic corrugation amplitude ($d$), one electronic ($d_e$), and the other elastic ($u$). The electronic component is caused by the spatial variation in the electronic density of states, which is normally observed with STM. For graphene on a substrate, this electronic height change ($d_e$) is known experimentally and theoretically to be at most 0.05 nm [25]. The elastic component is caused by the local elastic distortion for the freestanding graphene. In a related point, evidence for elastic distortions in graphite has been previously reported [40]. To illustrate, as the STM tip is brought into tunneling below a carbon atom, the freestanding graphene is attracted to the tip by the electrostatic image force and moves toward it (similar to our earlier discussion). The feedback circuitry retracts the tip until a stable equilibrium is achieved, as illustrated in the left side of Fig. 2(d). Next, as the tip scans away from the atom and toward the center of hexagon, the electrostatic force decreases. Consequently, the elastic restoring force causes the graphene sheet to retract away from the tip. The feedback circuit prompts the tip to chase the graphene until the new equilibrium configuration is reach as illustrated in the right side of Fig. 2(d). This repetitive, dynamic, and interactive process is in perfect registry with the electronic corrugation.

This study demonstrates that STM can be used to probe and induce a controllable strain on freestanding graphene membranes. These findings lay the groundwork for using local probes to study and manipulate the electrostatic/pseudo-magnetic/mechanical interactions in a model environment. For example, it may be possible to study local defects, such as vacancies or dopants, and their role in modifying the elastic and electronic properties of the membrane using STS.

In summary, we demonstrated that constant-current STS is a powerful tool to manipulate and study the atomic-scale properties of freestanding graphene. We also present atomic resolution STM images of freestanding graphene. The highly conductive yet flexible graphene membrane is shown to be attracted to the STM tip. The attractive force is found to periodically fluctuate from a maximum over the carbon atom to a minimum over the hole in the hexagonal structure giving rise to an extremely large atomic corrugation. As the graphene membrane moves towards the STM tip, an elastic restoring force builds up. From the induced strain we estimate the magnitude of the pseudo-magnetic field at the vicinity of the STM probe.

P.X. and P.T. acknowledge the financial support of the Office of Naval Research (ONR) under grant number N00014-10-1-0181 and the National Science Foundation (NSF) under grant number DMR-0855358. Y.Y. and L.B. thank the Office of Basic Energy Sciences, under contract ER-46612; NSF grants DMR-0701558, DMR-1066158 and MRI-0959124; ONR Grants N00014-11-1-0384, N00014-08-1-0915 and N00014-07-1-0825; and a Challenge grant from HPCMO of the U.S. Department of Defense. L.D. acknowledges financial support by the Taishan Overseas Scholar program, the National Natural Science Foundation of China (51172113), the Shandong Natural Science Foundation (JQ201118), the Research Corporation for Science Advancement, and NSF (DMR-0821159).

∗thibado@uark.edu

**Figure Captions**

FIG. 1 (color online). (a) SEM image of 2000-mesh copper grid with transferred graphene. (b) Constant-current scanning tunneling spectroscopy (CC-STS) data showing the vertical movement (*d*) of the STM tip versus applied bias voltage for three different setpoint currents. Bottom trace is graphene on copper data that has been offset for clarity. The inset shows the measured current as a function of applied bias voltage. (c) Schematic of the large-scale location of the STM tip below the freestanding graphene membrane with a low (left) and a high voltage (right).

FIG. 2 (color online). 6 nm × 6 nm filled-state (tip bias = +0.100 V, tunneling current = 1.0 nA) STM images of (a) graphene on copper. (b) Freestanding graphene. (c) Tip-induced graphene on HOPG graphite with a tunneling current of 0.3 nA. Upper Inset in (a-c): STM images cut from the center. Lower Inset in (a-c): Height cross-section line profiles taken from center. (d) Schematic illustrating the tip below a carbon atom (left) and below a hole in the honeycomb lattice (right).

FIG. 3 (color online). (a) Restoring force of graphene calculated from DFT versus normalized displacement, with a linear fit completed over the range from 0 to 0.1. (b) Atomic model used in the DFT calculations shown in a tilted view. (c) Force on a unit charge held 0.1 nm above the surface as it is moved parallel to the surface and along a line from the center of the honeycomb to the nearest carbon atom. A maximum force occurs over the carbon atom. (d) 5 nm radial average of the pseudo-magnetic field underneath the STM tip versus normalized displacement.

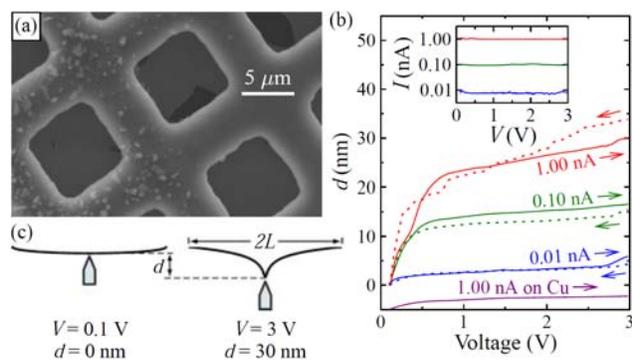

FIG. 1 by P. Xu et al.

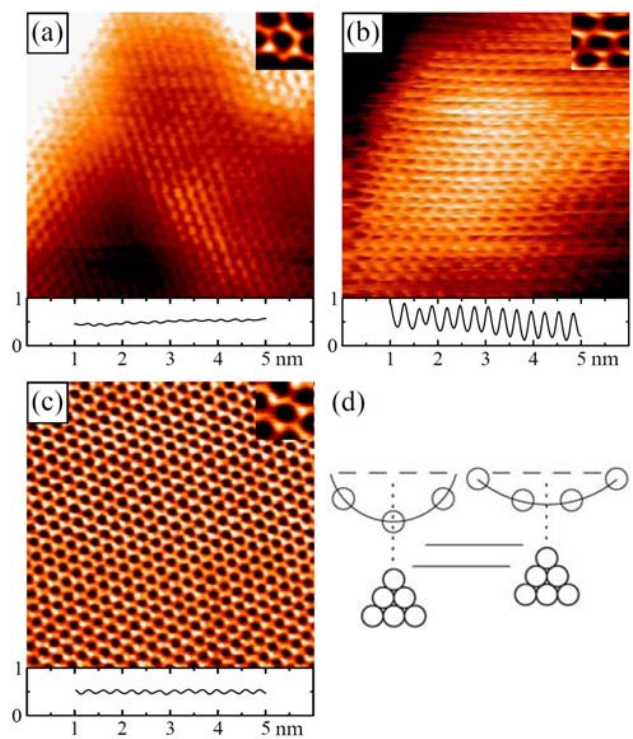

FIG. 2 by P. Xu et al.

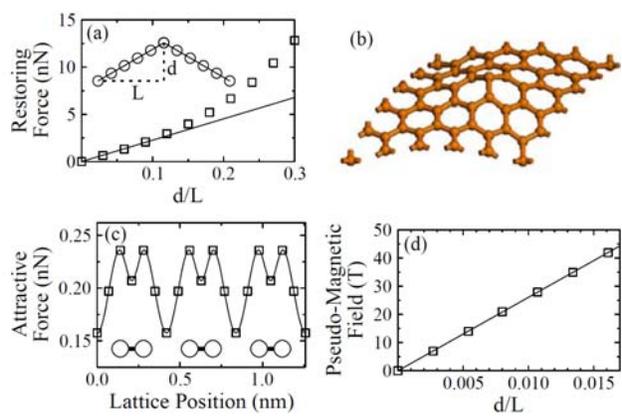

FIG. 3 by P. Xu et al.